\documentclass[preprint,prc,a4paper,tightenlines,nofootinbib]{revtex4}
\usepackage{psfrag}
\usepackage{epsfig}
 \usepackage{amsmath, amsthm, amssymb}
\newcommand{\la}{\langle}
\newcommand{\ra}{\rangle}
\newcommand{\p}{{\bf p}}
\newcommand{\PP}{{\bf P}}
\newcommand{\q}{{\bf q}}

\newcommand{\bsigma}{\mbox{\boldmath $\sigma$}}
\newcommand{\btheta}{\mbox{\boldmath $\Theta$}}
\newcommand{\bGamma}{\mbox{\boldmath $\Gamma$}}

\def\bi{\bibitem}
\newcommand{\beq}{\begin{equation}}
\newcommand{\eeq}{\end{equation}}
\newcommand{\beqa}{\begin{eqnarray}}
\newcommand{\eeqa}{\end{eqnarray}}

\newcommand{\be}{\begin{equation}}
\newcommand{\ee}{\end{equation}}
\newcommand{\nn}{\nonumber\\}
\newcommand{\eqn}[1]{\label{#1}}
\newcommand{\eq}[1]{Eq.~(\ref{#1})}
\newcommand{\eqs}[1]{Eqs.~(\ref{#1})}

\begin{document}
\preprint{FZJ-IKP-TH-2009-3}
\preprint{HISKP-TH-09/04}

\title{Gauge invariance in the presence of a cutoff
}
\author{A. N. Kvinikhidze$^{a}$}
\email{sasha_kvinikhidze@hotmail.com}
\author{B. Blankleider$^{b}$}
\email{boris.blankleider@flinders.edu.au}
\author{E.~Epelbaum$^{c,d}$}\email{e.epelbaum@fz-juelich.de}
\author{C.~Hanhart$^{c,e}$}\email{c.hanhart@fz-juelich.de}
\author{M.~Pav\'on Valderrama$^{c}$}\email{m.pavon.valderrama@fz-juelich.de}
\affiliation{$^{a}$A.~Ramzadze Mathematical Institute of Georgian Academy of
  Sciences, M.~Alexidze Str.~1, 380093 Tbilisi, Georgia \\ \\
$^{b}$Department of Physics,
 Flinders University, Bedford Park, South Australia 5042, Australia\\ \\
$^{c}$Forschungszentrum J\"ulich, Institut f\"ur Kernphysik (IKP-3) and
J\"ulich Center for Hadron Physics, 
             D-52425 J\"ulich, Germany\\ \\
$^{d}$Helmholtz-Institut f\"ur Strahlen- und Kernphysik (Theorie)
and Bethe Center for Theoretical Physics,
 Universit\"at Bonn, D-53115 Bonn, Germany\\ \\
$^{e}$Forschungszentrum J\"ulich, Institute for Advanced
Simulations, 
             D-52425 J\"ulich, Germany
}

\date{\today}

\begin{abstract}

  We use the method of gauging equations to construct the
  electromagnetic current operator for the two-nucleon system in a
  theory with a finite cutoff. The employed formulation ensures that
  the two-nucleon T-matrix and corresponding five-point function, in
  the cutoff theory, are identical to the ones formally defined by a
  reference theory without a cutoff. A feature of our approach is that
  it effectively introduces a cutoff into the reference theory in a
  way that maintains the long-range part of the exchange current
  operator; for applications to effective field theory (EFT), this
  property is usually sufficient to guarantee the predictive power of
  the resulting cutoff theory. In addition, our approach leads to
  Ward-Takahashi (WT) identities that are linear in the interactions.
  From the point of view of EFT's where such a WT identity is
  satisfied in the reference theory, this ensures that gauge
  invariance in the cutoff theory is maintained order by order in the
  expansion.
\end{abstract}

\pacs{XXX}

\maketitle


\section{Introduction}
Nuclear forces have been extensively studied over the past decade
within chiral effective field theory (EFT), see
Refs.~\cite{Bedaque:2002mn,Epelbaum:2005pn,Epelbaum:2008ga} for recent
review articles.  The chiral potential derived in this framework can
generally be split into its long- and short-range parts. The
long-range contributions are due to the exchange of one or more
Goldstone bosons [pions in the formulation based on the SU(2) chiral
symmetry of quantum chromodynamics (QCD)] and are strongly constrained by the spontaneously and
explicitly broken chiral symmetry of QCD. The long-range behavior of
the two-nucleon potential is furthermore independent of the
regularization procedure employed to evaluate the corresponding loop
integrals. The short-range contributions are usually parametrized in
terms of contact interactions in a most general way in order to ensure
that the results are model independent. It should be emphasized that
such contact interactions are not constrained by the chiral symmetry.
The a-priori unknown low-energy constants accompanying short-range
contact interactions have to be determined from two-nucleon scattering
data.

The potentials derived in chiral EFT, in general, are not well defined
at large momenta/short distances and cannot be used in a Schr\"odinger
equation without regularization. This is usually accounted for by
introducing a high-momentum/short-distance cutoff.\footnote{See
Refs.~\cite{Savage,Frederico:1999ps,irajdaniel,PavonValderrama:2005gu,PavonValderrama:2005wv,Yang:2007hb,Yang:2009kx}
for alternative renormalization schemes.} For applications to
electromagnetic reactions it is desirable to employ a mass-independent
regularization scheme such as, for example, dimensional regularization (DR), in
order to maintain gauge invariance at every stage of the calculation.
Given the nonperturbative nature of the nucleon-nucleon interaction at
low energy, it is not yet fully clear how to implement DR in this
context in the presence of long-range interactions (see however
Ref.~\cite{irajdaniel} for the first attempt along this line). It is
therefore important to clarify how cutoff regularization can be
carried out in the framework of chiral effective field theory without
destroying gauge invariance.

Such a procedure cannot be unique: the way to maintain gauge invariance 
will, in general, depend on how the  interaction without
photons is regularized. 
A lot of literature exists where formalisms are presented
to render the meson--baryon 
system gauge invariant~\cite{pins}.
An alternative method to regularize an interaction while preserving all
symmetries (including electromagnetic gauge invariance) was presented in
Ref.~\cite{Djukanovic:2004px} where a finite cutoff is introduced at the level
of the effective Lagrangian. For a different but related approach the reader
is referred to Ref.~\cite{Borasoy:2005zg}. 
In Refs.~\cite{dans,riska} a different, quite general construction for the $NN$ interaction is presented
that leads to gauge invariant currents; however, it is unclear how this recipe can be combined
with the power counting of an effective field theory.  
In this work we present a way to
introduce a cutoff while imposing the Ward-Takahashi (WT) identities for
various building blocks of the full reaction matrix elements.
The main advantages of our approach to the introduction of a cutoff are that
(i) it preserves the long-range part of the exchange current operator,
and (ii) it leads to the usual (linear in the interaction) WT identity for
regularized quantities, provided these identities are fulfilled by the same
quantities without regularization. The first feature is a necessary
requirement for any formulation based on effective field theory while
the second enables one to maintain \emph{exact} gauge invariance for
observables calculated at any fixed finite order in the EFT expansion.

At the same time, the predictive power of the scheme with cutoff is
expected to be the same as that of the corresponding theory without
cutoff. To see this, simply observe that the long-range parts of both
theories are identical by construction. In both theories the
short-range physics is parametrized by a series of contact
interactions. Their number is fixed by the symmetries of the theories
and thus is expected to be the same in both schemes, although the
actual values of the corresponding strength parameters will change.

In this paper we will introduce a new formalism to construct gauge
invariant amplitudes in the presence of non--perturbative interactions
using a cutoff which is not restricted to applications based on
effective field theories.  Consequently we will postpone all detailed
discussions intimately linked to effective field theories, like power
counting issues, to a subsequent paper.

The paper is organized as follows: in Sect.~\ref{sec:formal} we
formulate the problem, present our central results, and explain how to
calculate the amplitude for various electromagnetic transitions in the
two-nucleon sector.  In Sect.~\ref{sec:WT identity}, the central results of
our work are derived.  A brief summary is presented in
Sect.~\ref{sec:summary}.

\section{Formalism and important results} 
\label{sec:formal}

In this work we consider non-relativistic nucleons. It is,
therefore, preferable to define the reference operators in 
three-dimensional (3D) momentum space in such a way that gauge invariance
of the initial $\pi N \gamma$ Lagrangian is not destroyed. One possibility to
do this is discussed in Ref.~\cite{light-fr} where a  3D gauge-invariant
reduction in the light front formalism is carried out (the difference between
the light-front and usual time is not essential for this purpose).
A similar formalism is presented in Refs.~\cite{wallacephillips}. The basic
idea can be summarized as follows: gauge invariance is manifest for the usual
Green functions (vacuum expectation of the time ordered products of the field
operators) which fulfill the Ward-Takahashi identities but correspond to
operators acting in the space of four-momenta. A gauge invariant 3D reduction
can be achieved by equating the two-nucleon times which is equivalent to
integrating over the relative energy of the nucleons. This procedure does not
destroy the WT identity. There are also 
other possibilities of a gauge invariant 3D reduction such as 
e.~g.~on-mass-shell spectator reduction~\cite{riska,spect,gross}. For more
details on this topic the reader is referred to the original publications. 
We also emphasize that there exist various three-dimensional approaches to
derive nuclear potentials and current operators from meson-nucleon Lagrangians
which lead to a different form of the WT identity, see e.g.~\cite{Hyuga:1977cj} and 
Refs.~\cite{Friar:1992zz,Naus:1996eb,Naus:1997jg} for more
discussion on that issue. 
   
Consider the reference nucleon-nucleon $T$-matrix defined
through the formal expression 
\begin{equation}
T_{\rm ref}= V_{\rm ref} + V_{\rm ref} G_0(E) T_{\rm ref} \, .
\label{start}
\end{equation}
In this equation, the kernel $V_{\rm ref}$ refers to the potential
reduced to its three dimensional analog along the lines of
Ref.~\cite{light-fr}, and 
\be 
\langle \p'|G_0(E)|\p \rangle
= \delta(\p'-\p) \, \frac{M}{ME-\p^2+i\eta} \ , 
\ee
where $E$, $\p$ and $\p'$ 
denote the center-of-mass energy and the relative initial and final
momenta of the $NN$ pair, respectively, and $M$ is the nucleon mass.
%
Of central interest to this paper is the five-point function for two
nucleons interacting with a photon, $T_{\rm ref}^\mu$. In terms of diagrams, we shall
take $T_{\rm ref}^\mu$ to be the result of attaching the photon line everywhere inside of $T_{\rm ref}$ (but not to the external nucleon lines).\footnote{As discussed in Ref.~\cite{gaugingeq}, any contributions to  $T_{\rm ref}^\mu$ that cannot be obtained by attaching photons to $T_{\rm ref}$
are gauge invariant on their own, and can be added separately, as needed.} It can then be shown that \cite{gaugingeq}
\begin{equation}
T_{\rm ref}^\mu = (1+T_{\rm ref}G_0) V_{\rm ref}^\mu(1+G_0T_{\rm ref}) +
T_{\rm ref}G_0^\mu T_{\rm ref} \,,
\label{trefmu}
\end{equation}
where $V_{\rm ref}^\mu$  and $G_0^\mu$ denote the reference interaction
current (gauged potential) and the gauged two-nucleon propagator. 
In order to proceed, we need to assume that   $V_{\rm ref}^\mu$,
and hence $T_{\rm ref}^\mu$, obey the usual WT identities
\begin{subequations}
\begin{align}
q_\mu V_{\rm ref}^\mu&=\left[ \Gamma_0,V_{\rm ref}\right], \\[2mm]
q_\mu T_{\rm ref}^\mu&=\left[ \Gamma_0,T_{\rm ref}\right],
\end{align}
\end{subequations}
where $q^\mu$ denotes the four--momentum of an incoming photon and where
 we introduced $\Gamma_0$, defined in Eq.~(\ref{gamfree}),
to allow for a compact representation of the WT identity (see Sect.~\ref{sec:WT identityA}
for more details).
 
We define the T-matrix and the potential $V$ in the cutoff theory via 
\begin{subequations}\label{tdef}
\begin{align}
T &= V + V \, (G_0\Theta) \,  T, \label{tdefa} \\[1mm]
V &= V_{\rm ref} + V_{\rm ref} \, (G_0 \bar \Theta )  \, V, \label{tdefb}
\end{align}
\end{subequations}
where $\Theta$ ($\bar \Theta$) denotes the projector operator onto the space
of low (high) relative momenta with the usual properties $\Theta \bar \Theta =
\bar \Theta \Theta =0$ and $\Theta+\bar\Theta=1$. In particular, 
\beq
\la\p'|\Theta|\p\ra=\delta(\p'-\p)\theta(\Lambda-|{\bf p}|)
\eeq
 where $\Lambda$ is the cutoff momentum.
 Notice further that $[G_0,\Theta]=0$.  In distinction to
previous approaches to the same problem, we regard the cutoff as
part of the two-nucleon propagator. As a consequence, both $T$ and 
$V$ have an inverse. It is easy to see that $T$ equals $T_{\rm ref}$ exactly. 
Furthermore, it is  important to emphasize that the long-range parts
of $V$ and $V_{\rm ref}$ are identical by construction, since the term 
$V_{\rm ref} \, (G_0\bar \Theta )\, V$ is of a short range.  

We now need to gauge the above equations. One
finds in full analogy to Eq.~(\ref{trefmu}): 
\be
\label{Tmu}
T^\mu=(1+T G_0\Theta)V^\mu(1+ G_0\Theta T) +T(G_0\Theta)^\mu T  \,,
\ee
with the interaction current operator in the cutoff theory $V^\mu$ 
being defined via
\be
\label{Vmu}
V^\mu=(1+V G_0\bar\Theta)V_{\rm ref}^\mu (1+ G_0\bar\Theta V) +V\,
(G_0\bar\Theta)^\mu \, V \, .
\ee
We will show in Sect.~\ref{tequiv} that $T^\mu$ equals
$T^\mu_{\rm ref}$ exactly, and for properly choosen $(G_0\Theta )^\mu$ 
(see discussion in Sec.~\ref{sec:g0theta}), 
the current operators $V^\mu$ and  $V_{\rm ref}^\mu$ have the same 
long-range parts. Moreover, if the current operator $V^\mu_{\rm ref}$ and the
potential $V_{\rm ref}$ are related to each other via the usual WT identity, the same
holds true for the corresponding quantities  $V^\mu$ and
$V$ in the cutoff theory provided the completeness relation $(G_0\Theta)^\mu + (G_0\bar
\Theta)^\mu =G_0^\mu$ is satisfied. 


Thus, the actual problem reduces to constructing
$(G_0\Theta)^\mu$ which will be carried out in Sect.~\ref{sec:g0theta}.
As will be shown in this section, one possible choice is
\be 
 \la\p'|G_0^{-1}(G_0\Theta)^\mu G_0^{-1}|\p\ra = \frac{1}{2}\delta(\p_2'-\p_2)\,{\cal A}_1^\mu
  +(1\leftrightarrow  2) \eqn{reg-sim}
\ee
where
\begin{align} 
 {\cal A}_1^\mu &= \Gamma_1^\mu(\p_1',\p_1)
\left[ \theta(\Lambda-p')+ \theta(\Lambda-p)\right]\nn
 & +\left(0,\, \bGamma_1(\p',\p)\right)
 \frac{\theta(\Lambda-p')-\theta(\Lambda-p)}{{p}^{'2}-p^2}(E'_{cm}M-\p'^2+E_{cm}M-\p^2).
\eqn{onebodyco}
 \end{align}
In the above expression, $ \Gamma_1^\mu=\left(\Gamma_1^0,\, \bGamma_1\right)$ is the electromagnetic vertex of nucleon 1 [see \eq{Gamma^mu_i}], $\p_i$ ($\p_i'$) denotes the initial  (final) momentum of particle $i$, 
$p=|\p|$, $p'=|\p'|$, and $E_{cm}$ ($E'_{cm}$) is the energy of the two initial (final) nucleons in their own centre of mass system.
The modification of the free two-body current in the cutoff theory due to the
second term is crucial in
order to maintain the linear form of the resulting WT identity which leads
to \emph{exact} gauge invariance even in the case of approximate results
for $V$ and $V^\mu$ corresponding to a truncated iterative solution of
Eqs.~(\ref{tdef}) and (\ref{Vmu}). 

Note that in this work we only consider undressed nucleons and
the corresponding one-body currents~(\ref{onebodyco})
in order to explain the main ideas of our formalism.
Additional issues connected with
the gauge invariant dressing of the nucleons due to meson-nucleon interactions
and the proper treatment of relativistic corrections
will be discussed in a subsequent publication.

Before proving the results quoted above, we would like to
briefly remind the reader of the relations between $T^\mu$ and 
the amplitudes for various electromagnetic reactions. 
Clearly, the gauged $NN$ scattering amplitude $T^\mu$ contains the complete
information needed to describe such reactions (in the one-photon
approximation).
For instance,
in the case of the bremsstrahlung process one obtains 
\begin{eqnarray} \nonumber
\label{Abrems}
M^\mu(NN \to NN\gamma) &=&
G_0^{-1}(G_0TG_0)^\mu G_0^{-1} \\
&=& T^\mu+G_0^{-1}G_0^\mu T+TG_0^\mu G_0^{-1} \ ,
\end{eqnarray}
with $T^\mu$ 
defined in Eq.~(\ref{Tmu}).
\begin{figure}[t]
\begin{center}
\epsfig{file=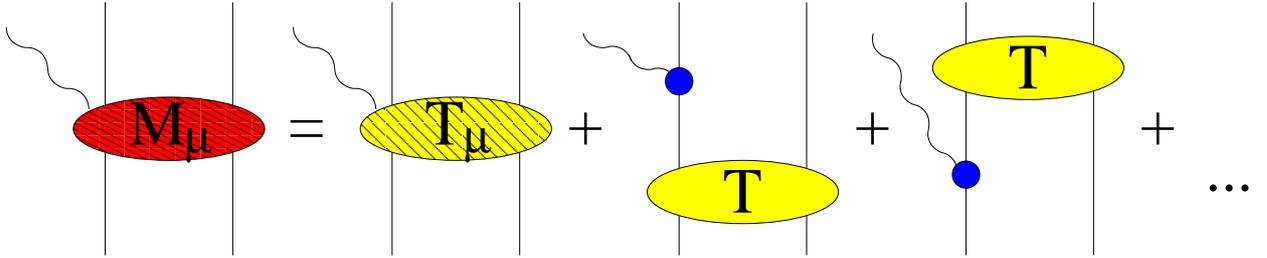,width=\textwidth}
\end{center}
\caption{\label{Bremspicture}
Graphical illustration for the full bremsstrahlungs amplitude.
Straight lines represent nucleons, wavy lines photons.
The diagrams where the photon is attached to
the other external nucleons are not shown.}
\end{figure}
The three terms in the last line of Eq.~(\ref{Abrems}) can be interpreted
diagrammatically in a straightforward manner (c.f. Fig.~\ref{Bremspicture}):
while the first term represents the coupling of the  photon
to the scattering matrix, the final two give the coupling to the
external legs. By construction, $T^\mu$ and $G_{0}^{\mu}$ obey
\begin{subequations}
\begin{align}
q_\mu T^\mu&=\left[ \Gamma_0,T\right], \\[1mm]
q_\mu G_{0}^{\mu}&=\left[ \Gamma_0,G_0\right],
\end{align}
\end{subequations}
where $\Gamma_0$ will be defined in Eq.~(\ref{gamfreemu}).
With this one immediately finds 
\begin{eqnarray}
q_\mu M^\mu(NN \to NN\gamma) = 0 \ ,
\end{eqnarray}
since all terms where $G_0^{-1}$ acts on an external state vanish.
Thus the formalism automatically produces conserved currents, which
should not come as a surprise since the conditions imposed by
the WT identities are stronger than just current conservation.

Eq.~(\ref{Abrems}) can be straightforwardly extended to
describe processes involving bound
states. To see this we introduce the vertex function
$\phi$ via
\begin{eqnarray}
\label{resid}
\langle {\bf p'} | T(E) | {\bf p} \rangle 
\stackrel{E \to E_B}{\longrightarrow}
\frac{\phi(\p')\,\phi^\dagger(\p)}{E-E_B}
\label{eq:T-res}
\end{eqnarray}
where $E_B$ is the bound-state energy.
Note, by construction $\phi$ fulfills the homogenous Lippmann-Schwinger
equation $\phi = V G_0 \Theta \phi$ and is related to the 
standard wave-function via $\psi_B(\p) = G_0\Theta \phi(\p)$.
Here we have already used that in the case of
non--relativistic kinematics, the bound state vertex
functions do not depend on the total momentum $P$.
With this the amplitude for the breakup reaction
may be written as $G_0^{-1}\left(G_0T\right)^\mu$
with the initial $T$--matrix evaluated at the bound state pole.
We thus have
\begin{eqnarray}
M^\mu(B\gamma\rightarrow NN) &=& 
 G_0^{-1}[G_0^\mu+G_0(1+T G_0\Theta)V^\mu 
G_0\Theta  + G_0 T(G_0\Theta)^\mu] \phi \nn
&=&[G_0^{-1}G_0^\mu+(1+T G_0\Theta)V^\mu 
G_0\Theta  + T(G_0\Theta)^\mu] \phi \, .
\end{eqnarray}
Similarly, the expression for
bound state form factors/transitions $B+\gamma\rightarrow B'$
reads
\begin{eqnarray}
\label{bs-cur}
\langle P'|J^\mu(0)|P\rangle =\phi_{P'}^\dagger[G_0\Theta V^\mu G_0\Theta
+ (G_0\Theta)^\mu]\phi_P \, .
\end{eqnarray}
We could have also derived this amplitude straightforwardly
from Eq.~(\ref{Tmu})
by  computing the residue of $T^{\mu}(E',E)$
at $E' = E_B'$ and $E = E_B$, where $E_B'$ and $E_B$ refer to the binding
energies of the final and inital states, respectively.

It may be useful to illustrate the numerical form of our operator equations by explicitly writing out Eq.~(\ref{bs-cur}) in terms of momentum-dependent variables. 
Using the explicit form for the single-nucleon current given in 
Eq.~(\ref{reg-sim}), and, for the sake of brevity, considering just the zeroth component
of the current, we have
\begin{eqnarray}
\label{eq:form-factor}
\langle P' | J^0(0) | P \rangle &=& 
\int \phi^\dagger_{P'}(\p')\,
\frac{d\p' \, \theta(\Lambda-p')}{E'_B-\p'^{\,2}/M} 
\, V^0(P',\p',P,\p) \, 
\frac{d\p \, \theta(\Lambda-p)}{E_B-\p^{2}/M}\,
\phi_P(\p)   
\\
&+&
\frac{i e_1}{2}\int \phi^\dagger_{P'}(\p')\,\frac{d\p \, [\theta(\Lambda-p ')
+ \theta(\Lambda-p) ]}{(E'_B-\p'^{\,2}/M)(E_B-\p^{2}/M)}\,
\phi_P(\p) + (1\leftrightarrow  2) \ .
\nonumber
\end{eqnarray}
Clearly, the first term on the right-hand side of this equation
describes the contribution from the exchange currents 
while the second and the third terms correspond to the impulse approximation (IA) current
(in which case $\p '=\p+\q/2$). 
We further emphasize that contrary to the more traditional approach
in which the cutoff is implemented in the potential rather than
in the two-nucleon propagator, the vertex function
$\phi(\p)$ extends up to infinite momenta.
The correct normalization of the vertex function $\phi(\p)$ can 
be read off from Eq.~(\ref{resid}) or, in case of energy-independent
potentials, from the usual wave function normalization condition
\begin{eqnarray}
\int d\p\,{\psi^\dagger_B(\p)} {\psi_B(\p)} =
\int d\p\,
\frac{\phi^\dagger(\p) \theta(\Lambda-p)}{E_B - \p^2/M}
\frac{{\phi}(\p) \theta(\Lambda-{p})}{E_B - \p^2/M} = 1 \, .
\end{eqnarray}

%
%

\section{Derivation of the central results}\label{sec:WT identity}

We now derive the results already quoted in the previous section. 
Our derivation is based on the WT identity for the 5-point functions. We therefore
begin with a brief discussion where the WT identity, and our notation for it,  are specified. 

\subsection{The WT identity for the 5-point function}
\label{sec:WT identityA}

\begin{figure}[t]
\begin{center}
\psfrag{TM}{\Huge{$T^\mu$}}
\psfrag{T}{\Huge{$T$}}
\psfrag{km}{\Huge{$q_\mu$}}
\psfrag{ei}{\Huge{$e_i$}}
\psfrag{kv}{\Huge{$\vec q$}}
\psfrag{pf0}{\large\hspace{-3mm}$(p_1',\ldots,p_{n'}')$}
\psfrag{pf1}{\large\hspace{-5mm}$(p_1',\ldots,p_{i}'-q,\ldots)$}
\psfrag{pi0}{\large\hspace{-3mm}$(p_1,\ldots,p_n)$}
\psfrag{pi1}{\large\hspace{-3mm}$(p_1,\ldots,p_i+q,\ldots)$}
\epsfig{file=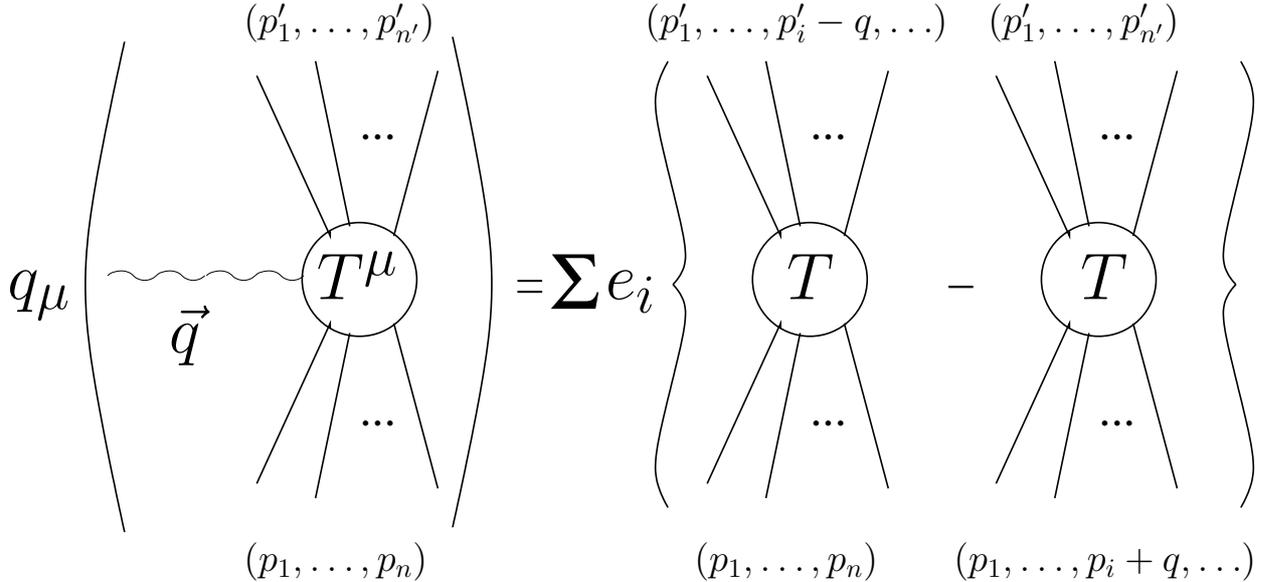,width=\textwidth}
\end{center}
\caption{Graphical illustration for the WT identity for $N$--point functions.
In our case we focus on $n=n'=2$ and therefore $N=5$.}
\end{figure}

A derivation of the WT identity for $N$-point functions can be found in most modern
textbooks on quantum field theory, see e.g.~Ref.~\cite{peskinschroeder}.  
The resulting expression reads
\begin{align}
q_\mu T^\mu&(q;p_1\ldots p_n;p'_1\ldots p'_{n'})\nonumber\\
&=\sum_i i e_i\left[
T\left(p_1\ldots p_n; p'_1\ldots , p'_i-q, \ldots\right)-T\left(p_1\ldots, p_i+q, \ldots; p'_1\ldots p'_{n'}\right)\right] \, . \label{wti_1}
\end{align}
In the case at hand, we have $n=n'=2$. 
It is more convenient for our purpose to rewrite the WT identity in 
operator form. Let 
$\Gamma_i^\mu$ denote the 
single-nucleon current operator of nucleon $i$, whose matrix element is extracted from the initial Lagrangian:\footnote{Here one could have equally well used the nonrelativistic limit of the Dirac-fermion current: 
\[
\Gamma^\mu_i(\p',\p)=ie_i\left(1,\, \frac{\p'+\p+i\bsigma_i\times \q}{2M}\right)\, .
\] 
}
\be
\Gamma^\mu_i(\p',\p)=ie_i\left(1,\, \frac{\p'+\p}{2M}\right)\, . \eqn{Gamma^mu_i}
\ee
Then the matrix element of the free two-nucleon current operator, $\Gamma_0^\mu\equiv G_0^{-1}G_0^\mu G_0^{-1}$, is given by
\begin{align}
\la \p'|\Gamma_0^\mu|\p\ra &=\delta(\p_2'-\p_2)\Gamma_1^\mu(\p'_1,\p_1) +\delta(\p_1'-\p_1)\Gamma_2^\mu(\p'_2,\p_2) . \label{gamfreemu}
\end{align}
The zeroth component of the free two-nucleon current is then (for simplicity of notation, we drop the zero superscript)
\begin{eqnarray}\label{gamfree}
\langle \p'|\Gamma_0|\p\rangle &=&ie_1\delta(\p_2'-\p_2)+ie_2\delta(\p_1' -\p_1)
\nn
&=& ie_1\delta(\p'-\p-\q/2)+ie_2\delta(\p'-\p+\q/2)
\end{eqnarray}
where $\q= \PP'- \PP$. Then the operator form of Eq.~(\ref{wti_1}) in the 
two nucleon case simply reads
\begin{equation}
q_\mu T^\mu=\left[ \Gamma_0,T\right] \ .
\label{WTI}
\end{equation}
This representation of the WT identity will prove very useful in subsequent
derivations.

\subsection{Transition current in the presence of a cutoff}
\label{tequiv}

It is straightforward to show that $T^\mu$ agrees identically with
 $T_{\rm ref}^{\mu}$ provided the following two conditions hold: 
(i) $V$ is a solution of Eq.~(\ref{tdefb}), and 
(ii) the free two-body current $(G_0 \Theta)^{\mu}$ fulfills the
completeness relation
\begin{equation}
(G_0\Theta)^\mu + (G_0\bar \Theta)^\mu = G_0^\mu \, . 
\label{gtcomp}
\end{equation}
Here, $G_0^\mu= G_0 \Gamma_0^\mu G_0$ denotes the free two-body
current before introducing the cutoff. 
The explicit form of $(G_0\Theta)^\mu$ is specified in Eqs.~(\ref{reg-sim})
and (\ref{onebodyco}) and derived in Sect.~\ref{sec:g0theta}.
With this one finds
\begin{eqnarray}
T^\mu&=&(1+T G_0\Theta)V^\mu(1+ G_0\Theta T) +T(G_0\Theta)^\mu T 
\nn
&=& (1 + T G_0\Theta)\left[ (1 + V G_0\bar\Theta) V_{\rm ref}^{\mu} 
(1+ G_0\bar\Theta V) + V (G_0\bar\Theta)^\mu V\right]
(1+ G_0\Theta T) + T(G_0\Theta)^\mu T
\nn
&=& (1 + T G_0\Theta + T G_0\bar\Theta) V_{\rm ref}^{\mu}
(1 + G_0\Theta T + G_0\bar\Theta T)
+ T(G_0\bar\Theta)^\mu T +T(G_0\Theta)^\mu T
\nn
&=& (1 + T G_0) V_{\rm ref}^{\mu} (1+ G_0 T) +T G_{0}^{\mu} T \nn
&=& T_{\rm ref}^{\mu}  \eqn{7} \ ,
 \end{eqnarray} 
where we have made use of \eq{tdefa}, \eq{Vmu}, and \eq{gtcomp}.
In practice, the equation defining $V$, \eq{tdefb}, 
cannot usually be solved exactly. 
However, what is important for applications based on effective field theory,
is that the long-range part of $V$ agree with the one of $V_{\rm ref}$.
In our formalism this holds by construction --- see also the discussion before
Eq.~(\ref{cond2}).
In effective field theory the transition potential $V_{\rm ref}$ and the
corresponding current operator $V_{\rm ref}^\mu$ are derived directly from an
underlying Lagrangian. We assume that this derivation is carried out in the
formulation which preserves the usual form of the WT identity:
$$
q_\mu V_{\rm ref}^\mu = [ \Gamma_0,V_{\rm ref}] \ . 
$$
We also need to demand that the free two-body current
$(G_0\Theta)^\mu$ obey the WT identity
\begin{equation}
q_\mu (G_0\Theta)^\mu = [ \Gamma_0,G_0\Theta] \ .
\label{g0tWTI}
\end{equation}
It is then straightforward to see that $V^\mu$ also obeys the usual WT identity; indeed,
\begin{eqnarray}
q_\mu V^\mu &=& 
(1 + V G_0 \bar\Theta) \, q_\mu V_{\rm ref}^{\,\mu}(1+ G_0 \bar\Theta V) +
V q_\mu (G_0\bar\Theta)^\mu V \nn
&=&
(1 + V G_0 \bar\Theta)[
\Gamma_0,V_{\rm ref}](1+ G_0\bar\Theta V) + 
V [ \Gamma_0, G_0\bar\Theta] V
\nn
&=&
(1+V G_0\bar\Theta)\Gamma_0 V  -
V \Gamma_0(1+ G_0\bar\Theta V) + 
V [ \Gamma_0,G_0\bar\Theta] V \nn
&=&
[ \Gamma_0,V]     \eqn{10} \,.
\end{eqnarray}
As a direct consequence, the five-point function $T^\mu$ will likewise satisfy the WT identity:
\begin{eqnarray}
q_\mu T^\mu =
[ \Gamma_0,T]\,.
\label{eq:WT identity-compact}
\end{eqnarray}
Eqs.~(\ref{Tmu}) and (\ref{10}) are all we need to
compute the five-point function in EFT with maximal predictive
power maintained.

\subsection{Construction of the free two-body currents}
\label{sec:g0theta}

So far we have not specified the explicit form of the operator $(G_0\Theta)^\mu$, apart
from requiring it to satisfy the WT identity,  \eq{g0tWTI}. In order to proceed, further  properties
of  $(G_0\Theta)^\mu$ need to be specified. In particular, we shall demand that:

\begin{enumerate}
\item The WT identity of Eq.~(\ref{g0tWTI}) must hold.

\item The integral given by the term $T(G_0\Theta)^\mu T$ in
  Eq.~(\ref{Tmu}) must involve only low relative momenta, such that the
  initial and final scattering amplitudes in Eq.~(\ref{Tmu}) are
  calculable in low-energy EFT. 
  Because of the completeness
  relation of Eq.~(\ref{gtcomp}), this condition also implies that
  the term $V(G_0\bar\Theta)^\mu V$ in Eq.~(\ref{Vmu}) is of
  short-range, i.e., $V^\mu$ has the same long-range/short-range
  decomposition as $V^\mu_{\rm ref}$. This can be achieved by
  demanding that $(G_0 \Theta)^\mu$ 
  provide a cutoff in a similar way to $G_0\Theta$, i.e.,
  $(G_0\Theta)^\mu=0$ if both initial and final relative momenta are
  above the cutoff parameter $\Lambda$; in particular, we shall demand that
  \be \langle
  \p'|G_0^{-1}(G_0\Theta)^\mu G_0^{-1}|\p\rangle =\delta(\p_2'-\p_2)[N^\mu
  \theta(\Lambda-p')+M^\mu\theta(\Lambda-p)]+(1\leftrightarrow 2)
\label{cond2}
  \ee
where $N^\mu$ and $M^\mu$ are yet to be determined.
\item
In the cutoff scheme,
the leading-order (LO) physical current in the {\it impulse approximation} [which is determined solely by the $(G_0\Theta)^\mu$
part of the current vertex]
 can differ from the one in the reference  
scheme only by terms of higher order.
This restriction has to do with the naturalness condition
and ensures that the effective interaction current $V^\mu$ does not violate 
power counting, as will be discussed in a subsequent paper.

\item 
The IA current $\la\p'|G_0^{-1}(G_0\Theta)^\mu G_0^{-1}|\p\rangle$ must be regular.

\item In the limit $\Lambda=\infty$, the IA current 
reduces to the usual expression of \eq{gamfreemu},
\be
\langle p'|G_0^{-1}(G_0\Theta)^\mu G_0^{-1}|p\rangle \ 
\rightarrow\ \delta(\p_2'-\p_2)\Gamma_1^\mu(\p'_1,\p_1) +\delta(\p_1'-\p_1)\Gamma_2^\mu(\p'_2,\p_2).
\ee

\item Time reversal invariance must hold:
\be\langle \p'|G_0^{-1}(G_0\Theta)^\mu
G_0^{-1}|\p\rangle =\langle \p|G_0^{-1}(G_0\Theta)^\mu G_0^{-1}|\p'\rangle  \,. \ee

 \end{enumerate}

The third restriction is imposed only on the IA part of the LO current to help
also the (WT identity based) relation between $V^\mu$ and $V$ to be the same
as the relation between $V^\mu_{\rm ref}$ and $V_{\rm ref}$. Here, 
the point is an ambiguity in solving WT identity for $V^\mu$ (WT identity can 
only be resolved for the longitudinal part of the current).
To see this, suppose that the interaction charge 
density in the reference scheme, $V^0_{\rm ref}$, were zero. 
If the relation between $V^\mu$ and $V$ is the same, $V^0$ should be zero as
well. This would not be achieved, if the LO physical current in the
{\it impulse approximation} in the cutoff scheme is different from
the one in the renormalization scheme in LO, because this difference
should be compensated by the interaction current, $V^0$.
The third  restriction can also be formulated as follows: $V^\mu$
should depend on the coupling constants composing $V$ in the same way
as $V^\mu_{\rm ref}$ depends on the coupling constants composing 
$V_{\rm ref}$. So we have six restrictions for $(G_0\Theta)^\mu$.

 It is interesting that the straightforward expression for
\begin{align}
 \langle \p'|G_0^{-1}&(G_0\Theta)^\mu G_0^{-1}|\p\rangle \nn
 &= \delta(\p_2'-\p_2)\Gamma_1^\mu(\p_1',\p_1)
 \frac{G^{-1}_0(p)\theta(\Lambda-p)-G^{-1}_0(p')\theta(\Lambda-p')}{E'-{p_1'}^2/2M-E+p_1^2/2M}+(1\leftrightarrow 2) \eqn{Gth-mu} 
\end{align}
is consistent with only five of the six restrictions. For example,
it cuts off high relative momenta and ensures that 
the term $V(G_0\bar\Theta)^\mu V$ in Eq.~(\ref{Vmu}) is of short-range
 so that $V^\mu$ has the same long-range/short-range 
decomposition as $V^\mu_{\rm ref}$  (in this case they have identical 
long-range parts). However, the vertex of \eq{Gth-mu} is singular in the case 
of inelastic scattering, where 
\be
E'_{cm}=E'-\PP'^{\,2}/4M\neq E-\PP^2/4M= E_{cm} \,,
\ee
and is therefore not consistent with restriction number 4.

Two examples of regular vertices are discussed in the Appendix, see 
Eqs.~(\ref{reg}) and (\ref{reg1}). To satisfy restriction number 6 (as well as the other five restrictions), we choose the
current vertex which is symmetric with respect to the interchange of the 
initial and final states as given previously in Eq.~(\ref{reg-sim}).
The derivation of this expression is carried out in the Appendix. It provides
useful insights into the gauge-invariant treatment of theories where a
cutoff is a necessary attribute in actual calculations (such as, for example, the NJL
model of Ref.\ \cite{bentz}). Gauge invariance can be maintained by a specific 
regularization, analogous to \eq{reg-sim}, of the integrals
corresponding to loops with an attached photon, in a close analogy to what 
is done in the present work. This regularization necessarily (and naturally) 
depends on the way the loops without photons are regularized.

It is also worth noting that the only terms which violate Galilean invariance in \eq{reg-sim}, namely, the combinations $\p_i'+\p_i=\PP'/2+\PP/2\pm(\p'+\p)$ contained in the single nucleon vertex functions $\Gamma^\mu_i$,
are the ones which enter the IA in the underlying theory.  The rest
depend only on the Galilei-invariant variables $\p', \p, E_{cm}$ and $E'_{cm}$.


Note finally that $\int d^3p\, [\theta(\Lambda-p')-\theta(\Lambda-p)]$
is of order $q$, and therefore one could think that the most unusual
part of the free two-nucleon current of \eq{reg-sim}, proportional to
$\theta(\Lambda-p')-\theta(\Lambda-p)$, may not contribute at lowest
order, since $q \ll\Lambda$. However, there is a compensating
enhancement from the denominator, since $\p'^{\,2}-\p^2$ is also of
order $q$.

 It is important to note that the
form of \eq{reg-sim} applies also to the practical case of a smooth
cutoff; that is, even if we replace the sharp cutoff $\theta$-functions by
smooth regulators, the WT identity will still be satisfied.  

\section{Summary}
\label{sec:summary}

In this paper we have shown how a finite cutoff can be implemented in
two-nucleon calculations without destroying the linear form of the WT identity and
without loosing any predictive power as compared to a mass independent
regularization like dimensional regularisation. 
In the latter case, we have
an $NN$ potential $V_{\rm ref}$ with long- and short-range parts, 
whose short-range couplings are determined from a fit to data.  
We also have an interaction current $V_{\rm ref}^\mu$ which is given by 
the initial Lagrangian and is restricted by gauge invariance. We
further assume that $V_{\rm ref}$ and $V_{\rm ref}^\mu$ fulfill the usual WT identity 
$q_\mu V_{\rm ref}^\mu=[ \Gamma_0,V_{\rm ref}]$ 
(see Sec.~\ref{sec:WT identityA}).
The physical $NN$ scattering amplitude and $NN$ currents are then
derived via \eq{start} and \eq{trefmu} respectively.

In the cutoff scheme we have an $NN$ potential $V$ with long- and 
short-range parts, whose short-range couplings are also determined
from a fit to the data.  We also have an
interaction current $V^\mu$ which, by construction, fulfills the
corresponding WT identity $q_\mu V^\mu=[ \Gamma_0,V]$. The
physical $NN$ scattering amplitude and $NN$ currents can be derived via
\eq{tdefa} and \eq{Tmu}, which represent the
cutoff versions of \eq{start} and \eq{trefmu}, respectively.
The currents in our cutoff scheme are conserved.

The difference between the approaches with the interactions $V_{\rm
  ref}$ and $V$ (and correspondingly with the currents $V_{\rm
  ref}^\mu$ and $V^\mu$) is in that $V_{\rm ref}$ is given directly by
the initial Lagrangian, whereas $V$ is related to $V_{\rm ref}$ in a
complicated way, see Eq.~(\ref{tdef}). This difference is not
important in the philosophy of EFT, because $V_{\rm ref}$ and $V$ have
the same long-range parts, and the short-range parts of $V_{\rm ref}$
and $V$ are anyway determined from experimental data.

The new technical element of our formulation is the 
single-nucleon current, $(G_0\Theta)^\mu$, which is
constructed to satisfy the WT identity in Eq.~(\ref{g0tWTI})
and is explicitly given by \eq{reg-sim}. It depends on the cutoff in a
specific way which is
consistent with gauge invariance.

Finally, we would like to comment on the relation of our interaction current
to the ones derived in Refs.~\cite{Nakamura:2006hc,Kvinikhidze:2007eu}. 
Although the current operators constructed in
\cite{Nakamura:2006hc,Kvinikhidze:2007eu} are sufficient to reproduce the
off-shell 5-point Green function (not only the physical
processes listed in Sect.~\ref{sec:formal}), they are related to the 
$NN$ potential $V$ via a {\it modified} WT identity which is {\it nonlinear} 
in $V$~\cite{Kvinikhidze:2007eu}. This is in strong contrast to the present
formulation where the usual linear WT identity is obtained. However, only on the basis
of a linear WT identity can gauge invariance be maintained order by order 
in some perturbative expansion.   

\section*{Acknowledgments}

We are greatful for discussions with U.-G. Mei\ss ner, K.~Nakayama,
D.R. Phillips, and D.O. Riska.  The work of E.E., M.P.V.~and
A.N.K.~was supported in parts by funds provided from the Helmholtz
Association to the young investigator group ``Few-Nucleon Systems in
Chiral Effective Field Theory'' (grant VH-NG-222) and through the
virtual institute ``Spin and strong QCD'' (grant VH-VI-231).  This
work was further supported by the DFG (SFB/TR 16 ``Subnuclear
Structure of Matter''), by the EU HadronPhysics2 project ``Study of
strongly interacting matter'' and by grant number GNSF/ST08/4-400
of the Georgian National Foundation.
\appendix

\def\theequation{\Alph{section}.\arabic{equation}}
\setcounter{equation}{0}
\section{IA current vertex}
\label{appb}
We shall define the momentum space matrix elements of the $\Theta$ operator as   
\begin{subequations}
\begin{align}
\la\p'|\Theta(P',P)|\p\ra &\equiv \Theta(\p'_1\p'_2,\p_1\p_2)=\delta(\p'-\p)\theta(\Lambda-p)\\
&=\delta(\p_1'-\p_1)\theta(\Lambda-p)=\delta(\p'_2-\p_2)\theta(\Lambda-p)
\end{align}
\end{subequations}
where $\p_1'+\p_2'=\p_1+\p_2$.
The regular solution of the WT identity of \eq{g0tWTI} for the free two-nucleon current
$(G_0\Theta)^\mu$, corresponds to the following gauged theta operator, $\Theta^\mu $:
\begin{subequations} \eqn{Theta^mu}
\begin{align}
\la\p'|\Theta^0(P',P)|\p\ra &=0\\
\la\p'|\btheta(P',P) |\p\ra
&=M\left[\delta(\p_2'-\p_2)\bGamma_1(\p',\p) -\delta(\p_1'-\p_1)\bGamma_2(\p',\p) \right]\nn
&\ \ \  \times \frac{\theta(\Lambda-p')-\theta(\Lambda-p)}{{p}^{'2}-p^2} . 
 \end{align}
\end{subequations}
It is then straightforward to check that $\Theta^\mu$ satisfies the usual WT identity:
 \begin{align}
 q_\mu \la \p'|\Theta^\mu&(P',P)|\p\ra = - (\p_1'+\p'_2-\p_1-\p_2)\cdot \la\p'|\btheta(P',P) |\p\ra\nn[2mm]
 =\ &- (\p_1'-\p_1)\cdot ie_1\delta({\bf p}_2'-{\bf p}_2)\frac{1}{2}({\bf p}'+{\bf p})\frac{\theta(\Lambda-p')-\theta(\Lambda-p)}{{p}^{'2}-p^2}\nn
&+(\p'_2-\p_2)\cdot ie_2\delta({\bf p}_1'-{\bf p}_1)\frac{1}{2}({\bf p}'+{\bf p})\frac{\theta(\Lambda-p')-\theta(\Lambda-p)}{p^{'2}-p^2}
\nn[2mm]
 =\ &- 2(\p'-\p)\cdot ie_1\delta({\bf p}_2'-{\bf p}_2)\frac{1}{2}({\bf p}'+{\bf p})\frac{\theta(\Lambda-p')-\theta(\Lambda-p)}{{p}^{'2}-p^2}\nn
&-2(\p'-\p)\cdot ie_2\delta({\bf p}_1'-{\bf p}_1)\frac{1}{2}({\bf p}'+{\bf p})\frac{\theta(\Lambda-p')-\theta(\Lambda-p)}{p^{'2}-p^2}
\nn[2mm]
 =\ &-ie_1\delta({\bf p}_2'-{\bf p}_2)[\theta(\Lambda-p')-\theta(\Lambda-p)]
-ie_2\delta({\bf p}_1'-{\bf p}_1)[\theta(\Lambda-p')-\theta(\Lambda-p)]\nn[3mm]
=\ &\ \ \ \, ie_1\Theta(\p'_1-\q,\p'_2,\p_1,\p_2)-\Theta(\p'_1,\p'_2,\p_1+\q,\p_2) ie_1\nn
&+ie_2\Theta(\p'_1,\p'_2-\q,\p_1,\p_2)-\Theta(\p'_1,\p'_2,\p_1,\p_2+\q) ie_2\, .
\end{align}

Before using \eqs{Theta^mu} to specify the IA current vertex, it is important to note that the use of the "product rule"
for gauging \cite{gaugingeq} gives
$[G_0\Theta]^\mu \equiv G_0^\mu \Theta+G_0\Theta^\mu$ and $[\Theta G_0]^\mu\equiv \Theta^\mu G_0 + \Theta G_0^\mu$, and therefore $[G_0\Theta]^\mu \ne [\Theta G_0]^\mu$ even though the operators $\Theta$ and $G_0$ commute,  $G_0\Theta = \Theta G_0$. It is, however, easy to check the obvious transversality of the difference, $q_\mu\left\{[\Theta G_0]^\mu- [G_0 \Theta]^\mu\right\}=0$. As expected, gauging alone can only determine the longitudinal part of the free two-nucleon current $(G_0\Theta)^\mu$ uniquely.

 Indeed, we can use either form to calculate the IA vertex current:
\begin{align} 
 \la\p'|&G_0^{-1}[G_0\Theta]^\mu G_0^{-1}|\p\ra=
 \la\p'|\left(\Gamma_0^\mu \Theta+\Theta^\mu G_0^{-1}\right)|\p\ra\nn
 &=\int d\p''\left[\delta(\p_2'-\p_2'')\Gamma_1^\mu(\p_1',\p_1'')
 +\delta(\p_1'-\p_1'')\Gamma_2^\mu(\p_2',\p_2'')\right]\delta(\p''-\p)\theta(\Lambda-p)\nn
 &+\left\{0,\, \left[\delta(\p_2'-\p_2)\bGamma_1(\p',\p) -\delta(\p_1'-\p_1)\bGamma_2(\p',\p) \right]\frac{\theta(\Lambda-p')-\theta(\Lambda-p)}{{p}^{'2}-p^2}(E_{cm}M-\p^2)\right\}\nn[3mm]
 &=\left[\delta(\p_2'-\p_2)\Gamma_1^\mu(\p_1',\p_1)
 +\delta(\p_1'-\p_1)\Gamma_2^\mu(\p_2',\p_2)\right]\theta(\Lambda-p)\nn
 &+\left\{0,\, \left[\delta(\p_2'-\p_2)\bGamma_1(\p',\p) -\delta(\p_1'-\p_1)\bGamma_2(\p',\p) \right]\frac{\theta(\Lambda-p')-\theta(\Lambda-p)}{{p}^{'2}-p^2}(E_{cm}M-\p^2)\right\}\nn[3mm]
&=\delta(\p_2'-\p_2)\left\{\Gamma_1^\mu(\p_1',\p_1) \theta(\Lambda-p)
 +\left[0,\, \bGamma_1(\p',\p)\right] \frac{\theta(\Lambda-p')-\theta(\Lambda-p)}{{p}^{'2}-p^2}(E_{cm}M-\p^2)\right\}\nn
 & \ \ \ + \ (1 \leftrightarrow 2), \eqn{reg}
 \end{align}
 or
  \begin{align} 
 \la\p'|&G_0^{-1}[\Theta G_0]^\mu G_0^{-1}|\p\ra=
 \la\p'|\left(\Theta \Gamma_0^\mu + G_0^{-1}\Theta^\mu\right)|\p\ra\nn
 &=\int d\p'' \delta(\p'-\p'')\theta(\Lambda-p'') \left[\delta(\p_2''-\p_2)\Gamma_1^\mu(\p_1'',\p_1)
 +\delta(\p_1''-\p_1)\Gamma_2^\mu(\p_2'',\p_2)\right]\nn
 &+\left\{0,\, (E'_{cm}M-\p'^2) \left[\delta(\p_2'-\p_2)\bGamma_1(\p',\p) -\delta(\p_1'-\p_1)\bGamma_2(\p',\p) \right]\frac{\theta(\Lambda-p')-\theta(\Lambda-p)}{{p}^{'2}-p^2}\right\}\nn[3mm]
 &=\left[\delta(\p_2'-\p_2)\Gamma_1^\mu(\p_1',\p_1)
 +\delta(\p_1'-\p_1)\Gamma_2^\mu(\p_2',\p_2)\right]\theta(\Lambda-p')\nn
 &+\left\{0,\, \left[\delta(\p_2'-\p_2)\bGamma_1(\p',\p) -\delta(\p_1'-\p_1)\bGamma_2(\p',\p) \right]\frac{\theta(\Lambda-p')-\theta(\Lambda-p)}{{p}^{'2}-p^2}(E'_{cm}M-\p'^2)\right\}\nn[3mm]
 &=\delta(\p_2'-\p_2)\left\{\Gamma_1^\mu(\p_1',\p_1) \theta(\Lambda-p')
 +\left[0,\, \bGamma_1(\p',\p)\right] \frac{\theta(\Lambda-p')-\theta(\Lambda-p)}{{p}^{'2}-p^2}(E'_{cm}M-\p^2)\right\}\nn
 & \ \ \ + \ (1 \leftrightarrow 2).  \eqn{reg1}
\end{align}
However, in order to satisfy time reversal invariance, it is better to use the symmetrized form $(G_0\Theta)^\mu = \frac{1}{2}\left\{[G_0\Theta]^\mu+[\Theta G_0]^\mu\right\}$:
  \begin{align} 
 \la\p'|&G_0^{-1} (G_0\Theta)^\mu G_0^{-1}|\p\ra
=\frac{1}{2}\delta(\p_2'-\p_2)\left\{ \rule{0pt}{15pt}\Gamma_1^\mu(\p_1',\p_1)
\left[ \theta(\Lambda-p)+ \theta(\Lambda-p')\right]\right.\nn
 & +\left.\left[0,\, \bGamma_1(\p',\p) 
 \frac{\theta(\Lambda-p')-\theta(\Lambda-p)}{{p}^{'2}-p^2}(E'_{cm}M-\p'^2+E_{cm}M-\p^2)\right]\right\}
 +(1\leftrightarrow  2) \eqn{reg-sim0}
 \end{align}
which is the form for the IA vertex current specified in \eq{reg-sim}.

It is instructive to rewrite \eq{reg} in a different form for the case where the single nucleon vertex current is of the form given in \eq{Gamma^mu_i}. We first write \eq{reg} as
\be 
 \la\p'|G_0^{-1}[G_0\Theta]^\mu G_0^{-1}|\p\ra = \delta(\p_2'-\p_2)\, A^\mu
  +(1\leftrightarrow  2) \eqn{reg2}
\ee
where
\begin{align} 
A^\mu &= \Gamma_1^\mu(\p_1',\p_1) \theta(\Lambda-p)
 +\left[0,\, \bGamma_1(\p',\p)\right]
 \frac{\theta(\Lambda-p')-\theta(\Lambda-p)}{{p'}^{2}-p^2}(E_{cm}M-\p^2).
 \end{align}
The zeroth and spatial components of $A^\mu$ can then be simplified as follows:
\begin{subequations}
\begin{eqnarray}
A_0/(ie_1)&=&\theta(\Lambda-p) \,,
\\
{\bf A}/(ie_1)
&=&\frac{1}{2M}(\p_1'+\p_1)\theta(\Lambda-p)
 +\frac{1}{2M}(\p'+\p)
 \frac{\theta(\Lambda-p')-\theta(\Lambda-p)}{{p'}^{2}-p^2}(E_{cm}M-\p^2),
\nn[5mm]
&=&\frac{1}{4M}(\PP'+\PP)\theta(\Lambda-p)
 +\frac{1}{2M}(\p'+\p)
 \frac{\theta(\Lambda-p')-\theta(\Lambda-p)}{{p'}^{2}-p^2} \,E_{cm}M \nn
&&
+\frac{1}{2M}(\p'+\p)
 \frac{\theta(\Lambda-p)\p'^2-\theta(\Lambda-p')\p^2}{{p'}^{2}-p^2}
\nn[5mm]
&=&\frac{1}{4M}(\PP'+\PP)\theta(\Lambda-p)+
\frac{1}{2M}(\p'+\p)\frac{\theta(\Lambda-p')-\theta(\Lambda-p)}{\p'^{\,2}-\p^2}(E_{cm}M+\frac{\p'^{\,2}}{2}+\frac{\p^2}{2})\nn
&& {}
+
\frac{1}{4M}(\p'+\p)[\theta(\Lambda-p)+\theta(\Lambda-p')]\,.
\end{eqnarray}
\end{subequations}

One can easily verify the WT identity for the operator in \eq{reg}. Expressing \eq{reg} as in \eq{reg2} we have
\begin{eqnarray}\label{TO-fix}
q_\mu A^\mu/(ie_1) &=&(P'-P)_\mu\left[\left(1,\frac{\p'_1+\p_1}{2M}\right)^\mu\theta(\Lambda-p')\right.\nn
&&
 \qquad \qquad 
 {} +\left.\left(0,\frac{\p'+\p}{2M}\right)^\mu 
\frac{\theta(\Lambda-p)-\theta(\Lambda-p)}{\p'^{\,2}-\p^2}(E_{cm}M-\p^{2})\right]
\nn[3mm]
&=& 
[(E'-E)-\p'^{\,2}_1/2M+\p^2_1/2M]\,\theta(\Lambda-p) \nn
&&
\qquad \qquad
{} -(\p'^{\,2}-\p^2)
\frac{\theta(\Lambda-p')-\theta(\Lambda-p)}{\p'^{\,2}-\p^2}(E_{cm}-\p^{2}/M)
\nn[3mm]
&=&
( E'-\p'^{\,2}_1/2M-\p'^{\,2}_2/2M ) \theta(\Lambda-p)-\theta(\Lambda-p')(E_{cm}-\p^{2}/M)
\nn[3mm]
&=&
G_0^{-1}(p')\theta(\Lambda-p)-\theta(\Lambda-p')G_0^{-1}(p)\nn[2mm]
&=&G_0^{-1}(p')\left[G_0(p)\theta(\Lambda-p)-G_0(p')\theta(\Lambda-p')\right]G_0^{-1}(p) \,.
\end{eqnarray}

\end{document}